# Spatial statistics of single-quantum detection

Jonathan F. Schonfeld

Harvard-Smithsonian Center for Astrophysics, Cambridge, Massachusetts 02140, USA

In a single-particle detection experiment, a wavefront impinges on a detector but observers only see a response at a single point. The extent of the wavefront becomes evident only in statistical accumulation of many independent detections, with probability given by the Born rule. Drawing on concepts from quantum optics, we analyze a simple model to reverse-engineer how this behavior can come about in terms of wave mechanics alone without a measurement axiom. The model detector consists of many molecules, each of which can be resonantly excited by the incoming particle and then emit a detection signature (e.g., localized flash of light). Different molecules have different resonant energies because local conditions (proximity of other molecules, Doppler shifts, etc.) vary. The detector is thus a quasi-continuum, and the incoming particle preferentially excites the molecule that it matches most closely in energy. (In actuality, molecules can be so numerous that many could closely match the incoming particle in energy; but in that case only one will be the first, and there will be nothing left for the others by the time the first match resonates and then emits. The model does not explicitly take into account the temporal advance of the particle wave packet through the detector medium.) The excited molecule can emit a detection signature, but that process competes with fluctuation-driven dephasing. We estimate the probability that a given molecule is resonantly excited, and we also estimate the probability that a detection signature is produced before being overwhelmed by fluctuations in the detector medium. The product of these two probabilities is proportional to the absolute-square of the incoming wavefunction at the molecule in question, i.e. the Born rule. We discuss possible ways to probe these mechanisms experimentally, and sanity-check the model with numbers from a neutron slit-diffraction experiment.

PACS number(s): 03.65.Ud, 42.50.Ct

## I. OVERVIEW

In this paper we draw on concepts from quantum optics in an attempt to understand the canonical behavior of macroscopic single-particle detectors: When a particle with a spread-out Schrodinger wavepacket approaches a detector, the result is a point signature (e.g. localized flash of light) rather than some sort of interference-pattern-shaped haze spread out along the wavefront. The haze does eventually appear, but only in the statistical accumulation of many independent point detections. The density of the haze is proportional to $|\psi|^2$, the absolute-square of the particle's wavefunction. (It is not *equal* to $|\psi|^2$ when normalized to the number of incoming particles; the proportionality constant reflects the detector's ability to react, and inefficiencies due to dissipation.) This has been documented with projectiles ranging from photons [1] to electrons [2] to neutrons [3] to Fullerenes [4] and beyond [5]. This general description also covers experiments (e.g., Stern-Gerlach) that use position-of-detection to measure non-position degrees of freedom.

In the literature, this behavior is typically asserted as an independent axiom that supplements Schrodinger's equation [6]. Without resorting to such an axiom, decoherence [7] explains how off-diagonal density-matrix elements of non-isolated systems relax to zero. But decoherence says nothing about why extended wavefronts produce point detections *before* statistical averaging. Without resorting to such an axiom, Ref. [8] analyzes the more general concept of wavefunction collapse in a model of a two-state system interacting with a bulk magnetic indicator, but that model doesn't address detector physics. Some authors refine the measurement axiom by replacing it with a quantifiable modification of non-relativistic Schrodinger physics. Ref. [9] describes a wavefunction-collapse-inducing modification of Schrodinger's equation; Ref. [10] describes an attempt to ascribe wavefunction collapse to gravity. The goal of the present paper is to reverse-engineer how point detection and the Born rule can come about from conventional non-relativistic wave mechanics without a measurement axiom or some sort of surrogate.

In order to do this, in Section II we posit a simple detector model based on a quantum optics paradigm. The model consists of many localized units – which we refer to for convenience as molecules without constraining generality – that can each be resonantly excited by the incoming particle and then emit a detection signature. (The underlying physics is reversible but its phenomenology is not because the phase space available for emission is much larger than the phase space occupied by the incoming particle – think of gas expanding into a previously empty volume.) Different molecules have different resonant energies because local conditions (proximity of other molecules, Doppler shifts, etc.) vary. The detector is thus a quasi-continuum [11], and the incoming particle preferentially excites the molecule that it matches most closely in energy. This is how detection is localized to a point. (In actuality, molecules can be so numerous that many can closely match the incoming particle in energy; but in that case only one will be the first, and there will be nothing left for the others by the time the first match resonates and then emits. The model does not explicitly take into account the temporal advance of the particle wave packet through the detector medium.)

In Section III, we will argue that the Born rule – which is logically independent of the statement that detection only occurs at a single point – comes about as follows: The excited molecule can produce a detection signature, but that process competes with fluctuation-driven dephasing (echoing the role of thermal fluctuations in decoherence theory). In Section III we will estimate the probability that a detection signature is produced before being overwhelmed by fluctuations in the detector medium (in the course of doing this we will need to make a conjecture about how best to describe the effect of fluctuations). We will also estimate the probability that a given molecule is resonantly excited in the first place. The product of these two estimates is the probability that a given molecule produces a detection signature. The Born rule results because this product is proportional to the absolute-square of the incoming wavefunction at the molecule in question. In this derivation, the Born rule is apparently not exact. Section IV discusses how one might observe deviations from exactness. This derivation of the Born rule also apparently involves details of the detector model, whereas in practice the rule is universally applicable. We conjecture that our model somehow captures the essential combinatorics of a more general statistical-mechanics universality argument.

Interestingly, our model implies that if a detection experiment were repeated with exactly the same incoming wavepacket and exactly the same detector molecules with exactly the same positions and thermal history, then detection would be observed (or not) at exactly the same point. It would thus appear that measurement-to-measurement randomness isn't intrinsic to quantum mechanics but instead reflects the practical impossibility of holding large numbers of molecules fixed for any experimentally meaningful period of time. However, it should be noted that in our model, molecular positions are assumed to have no intrinsic uncertainty. If we would describe molecular positions themselves with nontrivial wavefunctions, we might reach different conclusions about measurement repeatability.

Philosophically, we question whether Schrodinger's equation really needs a supplemental measurement axiom. After all, the equation on its own describes wave functions that can spread out into space or coalesce into bound dynamic entities. While physicists debate the meaning of wavefunctions, those coalesced entities, when sufficiently complex, can stumble on their own upon versions of thinking and acting, and eventually make their own versions of measurement. Something is very wrong if an axiom doesn't merely reflect what they will come to observe with or without our help.

To recapitulate, the remainder of this paper is organized as follows. Details of the model are introduced in Section II. Probabilities are estimated in Section III. Experimental tests and additional technical issues are discussed in Section IV. In an appendix we consider numerical values of model parameters for the neutron experiment in Ref. [3].

## II. MODEL

Following quantum optics convention [11], we posit the following normalized Hamiltonian for an incoming projectile interacting with detector molecules one at a time,

$$\hbar^{-1}H = \sum_n \omega_n |n\rangle\langle n| + \sum_{\mathbf{k}} \left[ \omega_{\mathbf{k}} |\mathbf{k}\rangle\langle\mathbf{k}| + \frac{\varepsilon}{L^{3/2}} \sum_n \left( e^{i\mathbf{k}\cdot\mathbf{r}_n} |\mathbf{k}\rangle\langle n| + e^{-i\mathbf{k}\cdot\mathbf{r}_n} |n\rangle\langle\mathbf{k}| \right) \right], \quad (1)$$

where $\mathbf{k}$ is projectile wavevector, $n$ enumerates detector molecules, $|n\rangle$ describes a single excited molecule (including absorbed or deflected projectile), $\mathbf{r}_n$ is the position of molecule $n$, $\varepsilon$ is a coupling constant, and $L$ is quantization volume. The set $\{\omega_n\}$ is the quasi-continuum indicated in Section I. Eq. (1) is expressed in terms of frequencies rather than energies because argumentation in Section III will involve competition between fundamental processes in time (numerical estimates in the appendix will be done more conventionally in terms of energy). Eq. (1) idealizes the incoming particle as a "scalar photon." This is obviously not a good idealization for all possible projectiles. We will try to apply it with some flexibility; see especially the appendix. Note that in Eq. (1), all the molecules interact with the incoming particle at the same time. As noted in Section I, we have made no provision for a wavefunction's gradual advance through the detector medium, so we will be able to reason only qualitatively about phenomena dominated by the first molecule to resonate.

An incoming wavepacket takes the form

$$|\text{in}\rangle = \left(\frac{2\pi}{L}\right)^{3/2} \sum_{\mathbf{k}} \varphi(\mathbf{k}) |\mathbf{k}\rangle, \quad (2)$$

where momentum- and position-space wavefunctions $\varphi$ and $\psi$ are related by

$$\psi(\mathbf{r}) = \left(\frac{2\pi}{L^2}\right)^{3/2} \sum_{\mathbf{k}} e^{-i\mathbf{k}\cdot\mathbf{r}} \varphi(\mathbf{k}). \quad (3)$$

If $|\text{in}\rangle$ is normalized to unity then, as usual,

$$\int |\psi(\mathbf{r})|^2 d^3r = \int |\varphi(\mathbf{k})|^2 d^3k = 1. \quad (4)$$

It follows that

$$\langle \text{in} | \hbar^{-1} H | n \rangle = \varepsilon \psi^*(\mathbf{r}_n). \quad (5)$$

So if the Hamiltonian is restricted to the space spanned by |in> and the $\{|n\rangle\}$, and the incoming wavepacket is assumed narrowband, then we can write

$$\hbar^{-1} H = \omega_{\text{in}} |\text{in}\rangle\langle\text{in}| + \sum_n \omega_n |n\rangle\langle n| + \varepsilon \sum_n (\psi^*(\mathbf{r}_n) |\text{in}\rangle\langle n| + \psi(\mathbf{r}_n) |n\rangle\langle\text{in}|). \quad (6)$$

It may seem overly restrictive to ignore the possibility of an excited molecule decaying back into a projectile state other than |in>. But we will assume that the detection-signature channel is so inexorable that back-decay doesn't really materialize.

If the separation between $\omega_{\text{in}}$ and a particular $\omega_n$ (presumably the first such $\omega_n$ encountered) is much smaller than the mean quasi-continuum spacing, then (6) reduces to a 2x2 matrix

$$\begin{pmatrix} \omega_{\text{in}} & \varepsilon \psi^*(\mathbf{r}_n) \\ \varepsilon \psi(\mathbf{r}_n) & \omega_n \end{pmatrix}. \quad (7)$$

To take into account subsequent emission of a strong detection signature, we modify (7) phenomenologically to include a decay width $\Gamma$,

$$\begin{pmatrix} \omega_{\text{in}} & \varepsilon \psi^*(\mathbf{r}_n) \\ \varepsilon \psi(\mathbf{r}_n) & \omega_n + i\hbar^{-1}\Gamma \end{pmatrix}. \quad (8)$$

For a resonant projectile and large $\Gamma$, i.e. for

$$|\omega_{\text{in}} - \omega_n| \ll \varepsilon |\psi(\mathbf{r}_n)| \ll \hbar^{-1}\Gamma, \quad (9)$$

then the two eigenvalues of Equation (8) reduce to

$$i\hbar^{-1}\Gamma, \quad i\varepsilon^2 |\psi(\mathbf{r}_n)|^2 / \hbar^{-1}\Gamma \quad (10)$$

(removing a common $\omega_{\text{in}}$). The corresponding eigenvectors are very close to (0,1) and (1,0), respectively. Since (1,0) is the initial condition in this projectile scenario, we conclude that nearly 100% of the projectile very slowly leaks into a strong detection signature. (Interestingly, Eq. (10) is the same eigenstructure as the neutrino mass matrix in the seesaw mechanism [12]).

In the appendix we discuss numerical values of the parameters in this section for the particular case of slow-neutron detection by gaseous boron-10 trifluoride.

## III. PROBABILITIES

It's easy to see that resonance (left-hand inequality in Eq. (9)) has probability roughly

$$P_{\text{resonance}} = 2\varepsilon |\psi(\mathbf{r}_n)| / \Omega, \quad (11)$$

where $\Omega$ is the bandwidth of the projectile wavepacket and we assume that a great many frequencies $\{\omega_n\}$ fall within this bandwidth but the bandwidth doesn't extend beyond the full range of $\{\omega_n\}$. It's encouraging that this scales with the absolute value of the projectile wavefunction, but it's not quadratic as expected from the Born rule.

(Eq. (11) should be viewed with caution because it has signal bandwidth do double duty as the size of a statistical ensemble.)

While the process governed by Eq. (10) is progressing, the resonant molecule is also subject to random perturbations from its surroundings and internally. We conjecture that, for the purpose of understanding the Born rule, these molecular fluctuations are best modeled as a Markov random walk through nearby internal energy levels. In this case energy-level probabilities conform to diffusion after many time steps simply as a consequence of the repetitive nature of random walk [13]. Accordingly, after time $\tau$ the probability distribution of this random walk in $\omega$ must take the general form

$$P(\omega - \omega_n, \tau) = \frac{1}{\sqrt{2\pi g\tau}} \exp -\frac{(\omega - \omega_n)^2}{2\pi g\tau} \quad (12)$$

for some phenomenological parameter $g$.

(This model is motivated by Ref. [14], which analyzes the density matrix of a damped quantum harmonic oscillator that begins in a pure state. In Ref. [14] the oscillator's wavefunction spreads to nearby energy levels, and the spreading of the wavefunction resembles diffusion in energy. We would have liked directly to conclude from this result that the same diffusion process characterizes energy-level *probabilities* in detector molecules. But we can't do this because a connection between wavefunction and probability is what we're trying to derive.)

Let us suppose that $|\omega-\omega_n|$ must be less than some rough threshold $G/2$ for the detection process described by the second eigenvalue in Eq. (10) to proceed (it seems intuitive that $G=O(\hbar^{-1}\Gamma)$). The probability of satisfying this inequality is roughly

$$GP(0, \tau) = \frac{G}{\sqrt{2\pi g\tau}}. \quad (13)$$

In order for detection to take place, this inequality must hold until a time equal to one over the second eigenvalue in Eq. (10). The probability of that happening is roughly

$$\frac{G}{\sqrt{2\pi g(\hbar^{-1}\Gamma/\varepsilon^2|\psi(\mathbf{r_n})|^2)}} = \frac{G\varepsilon|\psi(\mathbf{r_n})|}{\sqrt{2\pi g\hbar^{-1}\Gamma}} \approx \varepsilon|\psi(\mathbf{r_n})|\sqrt{\frac{\hbar^{-1}\Gamma}{2\pi g}}, \quad (14)$$

where we have taken the liberty of replacing $G$ by $\hbar^{-1}\Gamma$.

The total detection probability for molecule $n$ is the product of the right-hand-sides of Equations (11) and (14), i.e.

$$P_{\text{detection}} = \varepsilon^2|\psi(\mathbf{r_n})|^2 \sqrt{\frac{2\hbar^{-1}\Gamma}{\pi g\Omega^2}}$$

$$= |\psi(\mathbf{r_n})|^2 \left(\frac{\varepsilon^2}{\omega_{\text{in}}^2}\right)\left(\omega_{\text{in}}^2 \sqrt{\frac{2\hbar^{-1}\Gamma}{\pi g\Omega^2}}\right). \quad (15)$$

In Equation (15) we have factored the far-right expression into $|\psi|^2$, times an interaction volume that depends only on the incoming capture process, times a factor independent of the capture process that can be interpreted as quantum efficiency. In this form one recognizes Equation (15) as the Born rule, where, as usual, the proportionality constant multiplying $|\psi|^2$ reflects the detector's ability to react, and inefficiencies due to dissipation.

## IV. FURTHER CRITIQUE AND EXPERIMENTAL TESTS

We have introduced a microscopic model of a spatial detection process and analyzed it to reverse-engineer underlying reasons for localized detection and the Born rule. We have avoided conscious invocation of a measurement axiom. Nevertheless, we acknowledge that such an invocation could be implicit in the energy-diffusion reasoning embodied in Equation (12). Additionally, it is an open question whether all or even most common detection scenarios can actually be reduced in their essentials to this model.

The argumentation in this paper involves many approximations. Experiments that probe the model's underlying mechanisms could be designed to look for predictable deviations from these approximations. Here are some possibilities:

- If molecular positions and projectile parameters could be pre-specified with sufficient precision (consistent with the uncertainty principle), then perhaps detection could be made to take place at the same location, trial after trial, regardless of how $|\psi|$ depends on position.
- If dissipation could be suitably suppressed, then perhaps, following Eq. (11), detection probability could be made to be proportional to $|\psi|$ rather than $|\psi|^2$.
- If the incoming particle's bandwidth could be made sufficiently narrow, and the spread of molecular frequencies $\{\omega_n\}$ could be sufficiently biased as a function of position, then perhaps detection probability could be made to vanish within a spatially extended region, regardless of how $|\psi|$ depends on position.
- If the detector medium can be made sufficiently rarified, so that $\Omega$ < typical spacing between resonant energies, perhaps one can observe a crossover where the detector simply runs out of the statistics that drive the Born rule.
- Suppose the incoming wavefunction has transverse length scale $L$. Then, for fixed wavefunction depth (determined by bandwidth) the mean spacing between detector molecule energy levels scales like $L^{-2}$ but the mean wavefunction magnitude $|\psi|$ scales like the square root $L^{-1}$. So there can be crossover between values of $L$ for which the left-hand inequality in Eq. (9) can be satisfied by only one molecule and values for which the inequality is typically satisfied for more than one molecule. Perhaps this crossover can be observed.

## APPENDIX: NUMERICAL ESTIMATES OF MODEL PARAMETERS

In this appendix we map the foregoing model onto the neutron detection experiment of Ref. [3]. This provides a sanity-check for assumptions underlying at least the excitation half of our hypothetical detection mechanism. In that experiment, very slow neutrons are detected by $B^{10}F_3$ gas at standard temperature and pressure. The detection mechanism is

$$B^{10} + n \rightarrow (B^{11})^* \rightarrow Li^7 + \alpha. \quad (A1)$$

A zero-energy resonance is inserted to account for the anomalously large reaction cross section (see below). The excitation energy of the presumed intermediate state is

$$\Delta((B^{11})^*) = [m(B^{10}) + m(n) - m(B^{11})]c^2 = 12 \text{ MeV}. \quad (A2)$$

Other basic experimental parameters are

- Reaction cross section $\sigma \sim 4000$ bn = $4 \times 10^{-25}$ m$^2$ [15].
- Incoming wavenumber $k = 2\pi/(20\text{Å}) = \pi \times 10^9$ m$^{-1}$.
- Wavenumber bandwidth $\delta k = (2 \times 0.7\text{Å}/20\text{Å}) \times k = 7\% \, k$.
- Detector molecular number density $\rho \sim (3 \times 10^{-9}$ m$)^{-3}$ [16]. 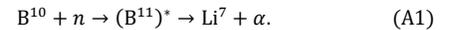
- Typical component of BF$_3$ velocity along projectile direction $v = [k_BT/m(BF_3)]^{-1/2} \sim 300$ m/s.
- Width of projectile wavefunction transverse to direction of motion at detector $w \sim 400 \, \mu$m = 0.4 mm.

These values imply the following:

- Total volume occupied by the incoming wavepacket $\sim w^2(2\pi/\delta k) \sim 4 \times 10^{-15}$ m$^3$.
- This determines the typical wavefunction magnitude $|\psi| \sim (4 \times 10^{-15}$ m$^3)^{-1/2} \sim 2 \times 10^7$ m$^{-3/2}$
- This volume also contains $\sim \rho \times (4 \times 10^{-15}$ m$^3) \sim 2 \times 10^{11}$ BF$_3$ molecules.
- The Doppler-driven spread in (B$^{11}$)* excitation energy across this sample $S \sim (v/c) \times 12$ MeV $\sim 12$ eV. (This is a lower bound because other effects could contribute to $S$.)
- So the typical separation between adjacent excitation levels $\delta \sim (12$ eV$)/(2 \times 10^{11}) = 6 \times 10^{-11}$ eV.
- The neutron kinetic energy $E = k^2\hbar^2/2m \sim 0.2$ meV.
- Neutron energy spread $\hbar\Omega \sim 2E\delta k/k \sim 3 \times 10^{-8}$ eV. This supports the assumption in the main part of this paper that $\delta \ll \hbar\Omega \ll S$.

To complete the work of this appendix, we try to estimate $\varepsilon$ from $\sigma$ by dimensional analysis. There is no unique way to do this, so we look at two alternative approaches. The first approach assumes the cross section arises from conventional perturbation theory in $\varepsilon$. The only cross-section-dimension combination of neutron mass, wavenumber and Planck's constant also quadratic in $\varepsilon$ is $\varepsilon^2 m^2/k^3\hbar^2$, implying $\varepsilon \sim 3 \times 10^{-6}$ s$^{-1}$m$^{3/2}$. Alternatively, we can assume that the large neutron-capture cross section indicates a near-zero-energy resonance and therefore must approximately scale like $k^{-2}$ [17]. But this scaling already has the dimensions of cross section, so this approach cannot determine $\varepsilon$ within this paper's parameter-poor model. The perturbative estimate implies $\hbar\varepsilon|\psi| \sim 4 \times 10^{-14}$ eV $\ll \delta$. This supports the quasi-continuum assumption, according to which the incoming particle can distinguish between different detector molecules by resonant frequency, and in this case at most one molecule can be excited (no need for a first-to-resonate assumption). However, this estimate is highly suspect since cross section scaling like $k^{-3}$ at a zero-energy resonance is not physically reasonable.

**References**

1. R. Aspden, M. Padgett, and G. Spalding, Am. J. Phys. **84**, 671 (2016).
2. A. Tonomura et al, Am. J. Phys. **57**, 117 (1989).
3. A. Zeilinger et al, Rev. Mod. Phys. **60**, 1067 (1988).
4. L. Hackermuller et al, Nature **427**, 711 (2004).
5. S. Gerlich et al, Nature Communications **2**, 263 (2011).
6. R. Feynman, R. Leighton, and M. Sands, *The Feynman Lectures on Physics* (Addison-Wesley, Reading, Mass. 1965), Vol. III.
7. M. Schlosshauer, Rev. Mod. Phys. **76**, 1267 (2004).
8. A. Allahverdyan, R. Balian, and T. Nieuwenhuizen, Phys. Rep. **525**, 1 (2013).
9. A. Bassi, J. Phys. Conference Series **701**, 012012 (2016).
10. R. Penrose, Gen. Relativ. Gravit. **28**, 581 (1996).
11. E. Kyrola and J. Eberly, J. Chem. Phys. **82**, 1841 (1985).
12. S. King, Rept. Prog. Phys. **67**, 107 (2004).
13. S. Chandrasekhar, Rev. Mod. Phys. **15**, 1 (1943).
14. B. Zel'dovich, A. Perelomov, and V. Popov, Sov. Phys. JETP **28**, 308 (1969).
15. V. Sears, Neutron News **3**, 26 (1992).
16. Honeywell Corporation, "Honeywell Boron Trifluoride Technical Information."
17. J. Taylor, *Scattering Theory* (John Wiley & Sons, New York 1972).